\documentclass{aastex62}
\UseRawInputEncoding
\submitjournal{ApJ}
\def \kepler{{\slshape Kepler}}

\begin{document}
{\affiliation{Department of Astronomy, University of Florida, 211 Bryant Space Science Center, Gainesville, FL, 32611, USA}}
\title{Signatures of impact-driven atmospheric loss in large ensembles of exoplanets}

\author[0000-0002-9916-3517]{Quadry Chance}
\affiliation{Department of Astronomy, University of Florida, 211 Bryant Space Science Center, Gainesville, FL, 32611, USA}
\author[0000-0002-3247-5081]{Sarah Ballard}
\affiliation{Department of Astronomy, University of Florida, 211 Bryant Space Science Center, Gainesville, FL, 32611, USA}
\author[0000-0002-3481-9052]{Keivan Stassun}
\affiliation{Department of Physics \& Astronomy, Vanderbilt University, Nashville, TN 37235, USA}

\begin{abstract}
The results of large-scale exoplanet transit surveys indicate that the distribution of small planet radii is likely sculpted by atmospheric loss. Several possible physical mechanisms exist for this loss of primordial atmospheres, each of which produces a different set of observational signatures. In this study, we investigate the impact-driven mode of atmosphere loss via N-body simulations. We compare the results from giant impacts, at a demographic level, to results from another commonly-invoked method of atmosphere loss: photoevaporation. Applying two different loss prescriptions to the same sets of planets, we then examine the resulting distributions of planets with retained primordial atmospheres. As a result of this comparison, we identify two new pathways toward discerning the dominant atmospheric loss mechanism at work.  Both of these pathways involve using transit multiplicity as a diagnostic, in examining the results of follow-up atmospheric and radial velocity surveys.
\end{abstract}

\section{Introduction}
The study of exoplanetary atmospheres, even in its first decades, is characterized by incredible diversity and complexity. Attempts to link the phenomenology of atmospheres to the dominant underlying processes have been correspondingly inventive. 
While placing individual planets under a microscope is a critical exercise, so too is studying large-scale demographic patterns among samples of many atmospheres. The study of exoplanetary demographics was transformed by NASA's \kepler{} Mission. The discoveries of of thousands of super-Earths with radii between $1-4 R_{\oplus}$, the most common type in the Milky Way \citep{Fressin13,Howard2013,Dressing15,Bonomo2019}, have crucially informed theories of planet formation and evolution, including the processes that shape their atmospheres. Many of the most detailed studies atmospheric demographics await missions like the James Webb Space Telescope (hereafter JWST, Gardner et al 2006). However, with sample sizes of several planets to dozens, many studies (\citep{Sing2016, Crossfield2017}, and \citep{Guo17} et al. among others) are beginning to illuminate patterns among the rich complexity of exoplanetary atmospheres. 

 One of the defining features of Super-Earths, as they are currently understood, is the bimodal radius distribution of small, short period exoplanets: there exists a gap in their radii at around $1.5-2 R_{\oplus}$ \citep{Fulton2017}. As our constraints of the planets' stellar properties have improved, our understanding of the radius gap has grown \citep{Fulton18}, including whether planets reside in the gap proper. Models have predicted the presence of such a gap before observations of a large enough population revealed its presence \citep{Ciardi2013, Owen2013}. Several possible mechanisms have emerged to explain this feature. Among them are photoevaporation of the atmospheres by stellar irradiation \citep{Owen2013,Pu2015a, Berger20}, loss due to the heating of the atmosphere by the newly-formed hot planetary cores (or ``core-powered mass loss", \citealt{Ginzburg2018,Owen2016}), a population of primordially bare planets \citep{Lopez2018}, and atmospheric loss due to giant impacts with debris or other planets \citep{Biersteker2019}.
 
 The known sample of thousands of exoplanets is presumably shaped by a wide variety of physics, both imprinted from formation and ongoing-- however, there are key ways in which predictions diverge for different physics \citep{Owen2017}.  All three methods can produce the "Fulton gap" between the subset of planets with retained primordial atmospheres and those without. But the existence of adjacent planets with extremely different densities, for example, cannot be explained only by a mechanism that \textit{only} ever depends the planets' distance from the host star \citep{Bonomo2019,Inamdar2016, Carter2012}. The observational consequences of different processes include (among other properties) the envelope fraction, bulk density, and atmospheric composition of exoplanets. On a large-scale, the underlying dominant processes will shape exoplanet demographics as a whole in ways we can model and predict \citep{Dawson2016,MacDonald2020}. 
 
 In this paper, we investigate the giant impact mechanism for atmospheric loss from a demographic standpoint. Using a suite of N-body simulations to track impacts during formation, we examine the properties of the resulting planets. \cite{Biersteker2019} have shown that giant impacts have an erosive effect on the primordial atmospheres of terrestrial planets, with complete loss possible in some cases. This effect would be \textit{a priori} stochastic in nature: a single impactor can significantly alter the core/envelope mass ratio, resulting in a large, observable density difference even between adjacent planets. For the same suite of planets, we also apply the competing photoevaporation model, in order to compare the resulting populations of planets in a head-to-head fashion. We will not, however, be considering the effect of core-powered mass loss
 
This manuscript is organized as follows. In Section \ref{sec:methods}, we describe the extant N body simulations we employ for this study. We detail how we translate the simulation outcomes to observable planet properties, for comparison to the results of large-scale transit survey (in particular, NASA's \textit{Kepler} Mission). We detail our set of assumptions for assessing whether atmospheric loss has occurred for a given planet, dependent on whether giant impacts or photoevaporation is dominant. In Section \ref{sec:analysis}, we describe the ways in which these two processes sculpt the resulting population of planets from our simulations, highlighting their differences.  In Section 4, we explore the efficacy of different observational tools to differentiate between these two hypotheses, based upon our findings. In Section 5, we summarize our findings and conclude. 

 \section{Methods}
 \label{sec:methods}
 \subsection{Details of N-body simulations}

To interpret the effects of giant impacts on the primordial atmospheres on a large set of exoplanets, we employ a suite of late-stage planet formation N-body simulations. A set of simulations published in \cite{Dawson16}, with details included therein, are ideal for our experiment. They span the relevant timescale for planet formation, nominally 10 Myr, so that we can track each impact as planets accrete. They also demonstrably recover key observable features among mature planetary systems. \cite{Dawson16} (hereafter D16) simulated late-stage planet formation around stars with mass of 1 $M_{\odot}$ over a range of solid and gas disk surface densities. They assumed a common surface density power law ($\alpha=3/2$, per the minimum mass solar nebula), over typical timescales of 10 Myr.  For detailed description of the simulation initial conditions, we refer the reader to D16. They went on to investigate the links between the initial conditions of these simulations and the resulting dynamical conditions of the resulting planets.  We use these simulations to investigate a new, but possibly linked, phenomenon: the role of giant impacts in sculpting atmospheric loss. These simulations are useful toward that end, firstly because we can trace every impact of two bodies and their corresponding masses. Secondly, the process of accruing a primordial atmosphere happens contemporaneously with planet assembly. Our analysis of the output differs from D16 in that we do not apply any prescription for gas accretion; we make the simplifying assumption that all planets left at the gas disk's dissipation acquire a primordial H/He atmosphere.

We employ a key finding in that work for this study as well, which is that a combination of disk initial conditions is necessary to recover the properties of observed \textit{Kepler} exoplanetary systems as a whole. No single disk model alone in that work reproduced the \textit{Kepler} distributions in transiting planets per system, transit duration ratios of adjacent planets, and period ratios of adjacent planets. The necessity for a mixture model of initial disk conditions is attributable to the wide range of ``dynamical temperatures" \citep{Tremaine2013} among planetary systems. Simulations with different initial conditions in solid surface density and gas surface density, result in variation in average dynamical temperature (see also \citealt{Moriarty16}, who reached a similar conclusion). Exoplanetary systems with wide orbit spacings, high eccentricities, and high mutual inclinations are canonically ``dynamically hotter"; the ensembles that produce these simulations are denoted \texttt{Ed$10^{4}$} in D16 and are initialized with a greater degree of gas depletion and therefore less strongly damped. In contrast, systems with tighter orbit spacings, lower orbital eccentricities, and lower mutual inclinations are ``dynamically colder"; these systems originate with start with similar disks with less gas depletion and are denoted \texttt{Ed$10^{2}$} in that work. In the recent exoplanet literature, these latter dynamically cool and densely populated ``Systems of Tightly-packed Inner Planets" are denoted ``STIPS'' \citep{Volk2015}; we employ this latter term throughout. 

\subsection{Assembly of Planetary System Sample}
\label{sec:assembly}

In the original D16 study, the authors determined that matching the ensemble of \textit{Kepler} planets to the output planets from disk simulations required a combination of disk types. D16 explored a large range of intial surface densities, inclination  and eccentricity distributions, and spacings of the planet embryos. \citep{Moriarty2016}, who varied the surface density profile as well as total mass available, came to similar conclusions. The necessity for this mixture model is the subject of active discussion. Initial studies of Kepler's multi-transiting planet systems \citep{Lissauer11, Johansen2012} argued for a so-called ``Kepler dichotomy", but later studies showed that modifications to key assumptions about sensitivity of the Kepler observations to multiple transiting planets \citep{Zink19} or to the underlying relationship between number of planets and mutual inclination \citep{Zhu18, He20, Millholland21} made a "single mode" model tractable. Because the mixture model, as a phenomenological descriptor,  recovers important observables \citep{He19}, we adopt it as a heuristic for this study. In this paper, singly-transiting systems or singles refer to systems that only have one transiting planet while multiple-transiting or multis refers to systems with multiple transiting planets.

There exist multiple estimates for the underlying fraction of planetary systems that are drawn from each population: the STIPS are heavily over-represented in transit surveys. This is due to the larger number of planets overall, in addition to the likelier transit probability at shorter orbital periods. The underlying rate of compact systems of multiple planets is estimated at 5\% of the total \citep{Lissauer11, Contreras18}, though they can make up half or more of detected planets \citep{He19, Ballard16}. \citep{Moriarty16} found that the underlying rate of compact multiples around FGK stars is half that of the rate around M dwarfs; the latter lies between 10-20\% with 1$\sigma$confidence \citep{Muirhead15, Ballard19}. For this study, we adopt an underlying compact multiple rate of 7\% for Sunlike Stars; that is, 7\% of planetary systems are drawn from the ensemble \texttt{Ed$10^{2}$} (which produces most STIPS), while 93\% are drawn from \texttt{Ed$10^{4}$}. To create a synthetic population of planets to realistically compare to a Kepler-like survey, we use this 93\% ``dynamically hot"/7\% ``dynamically cool" partition to draw ~150,000 stars with randomly-oriented ecliptic angles. The systems are drawn uniformly from the set of 80 \texttt{Ed$10^{2}$} and 80 \texttt{Ed$10^{4}$} systems.

\subsection{Assigning Radii to Planets}

We employ two prescriptions for atmospheric loss with each set of simulations: one in which photoevaporation is the dominant process and one in which giant impacts are dominant. The choice of which is dominant determines how we translate a planetary mass output by the n-body code used in D16,\textsc{MERCURY6} \citep{chambers12} to a radius of that planet today.  We assume that, all else being equal, a planet will accrete and retain a primordial atmosphere from the disk unless that process is inhibited or halted; we vary only the assumption about how that loss might occur. In reality, the gas accretion rate of a planet is sensitive to the mass of the planet. In practice, the 1 My timescale for accretion is more than sufficient for the majority of our simulated planets to accrete primordial atmospheres \citep{Ikoma2012}.  In-situ accretion of H-dominated atmospheres is thought to result in envelope mass fraction from $10^{-2}$ to $10^{-1}$ for planets with masses $< 10M_{\oplus}$, with the envelope fraction increasing with the core mass \citep{Ikoma2012}: for this work, we have assumed an initial envelope mass fraction of 5\%. This assumption will increase our planet radii by roughly 10\%, which corresponds to a ~ 20\% increase in the transiting planet yield all else being equal. Each process for atmospheric sculpting, photoevaporation and giant impacts, has a set of necessary conditions under which the planet loses this primordial H/He atmosphere. 

To isolate the effects of each process, we assume first that the two processes are mutually exclusive. For planets that retain their primordial atmospheres, we use the mass-radius models of \cite{Lopez2013} for the appropriate remaining gas fraction and equilibrium temperature. For these planets, their radius is defined at 20 mbar. For planets that lose their atmospheres, we use the Earth-like rocky composition models of \cite{Zeng2019}. In Subsection \ref{sec:atm_loss_impact}, we first consider giant impacts and the conditions under which planets retain their primordial atmospheres at the end of the simulations. In subsection \ref{sec:atm_loss_photo}, we turn to photoevaporation to ask the same. 

\subsubsection{Loss Driven by Giant Impacts}
\label{sec:atm_loss_impact}

We employ the formalism of \citep{Biersteker2019} to determine the erosive effect of giant impacts on the primordial atmospheres of terrestrial planets. There are two primary erosive mechanisms at work when planet embryos collide. The first is atmospheric displacement from the impact-generated shockwave. \added{Detailed in \cite{Schlichting18}}, eroding the atmosphere completely with only a few impacts requires impactors with radii comparable to the target embryo. The second mechanism is loss due to the thermal energy deposited into the planet by the impact. For planets fresh with the heat of formation, the energy required to heat the base of the envelope past this threshold may be very low, leading to complete envelope stripping from lower energy impacts. The energy required to remove an envelope this way depends on the energy budget of the planet separated into the thermal energy in its core and envelope. In the core-dominated regime, the envelope is stripped when the Bondi radius is equal to the core radius, so that gas molecules are no longer gravitationally bound to the planet.  

We consider here the most conservative scenario for atmosphere retention- in which the core thermal energy can be neglected in favor of thermal energy delivered by the impactor- the entire envelope will be ejected when $E_{imp} + E_{env}$ = 0 where $E_{imp}$ is (from \citealt{Biersteker2019} Equation 18)and $v_{imp} = v_{esc}$, and $E_{env}$ is approximated as:
\begin{equation}
E_{env} = \eta GM_{env} \frac{M_{c}}{R_{c}}
\end{equation}
where $\eta$ is the efficiency of energy transfer from the impactor to the core and envelope of the target. 
Impacts can happen at speeds reaching a few times the lower limit of $v_{esc}$. Since the impact velocities were not tracked in the original D16 simulations, we ignore the details of each collision make the conservative assumption that they all happen at  $v_{esc}$.
In the limit where all of the impact energy goes into heating the core and envelope ($\eta$ = 1) and an envelope mass of $M_{env} = 0.05M_{c}$, the condition required to eject an envelope is: 

\begin{equation}
M_{imp}G\frac{M_{p}}{R_{c}} + GM_{env}\frac{M_{c}}{R_{c}} = 0
\end{equation} 
Rearranging this to get an expression in terms of impactor mass:
\begin{equation}
M_{imp} \frac{1-E_{env}}{\eta}M_{env} =\frac{f}{\eta} M_{c}
\end{equation}

\begin{figure}
\includegraphics[width=12cm]{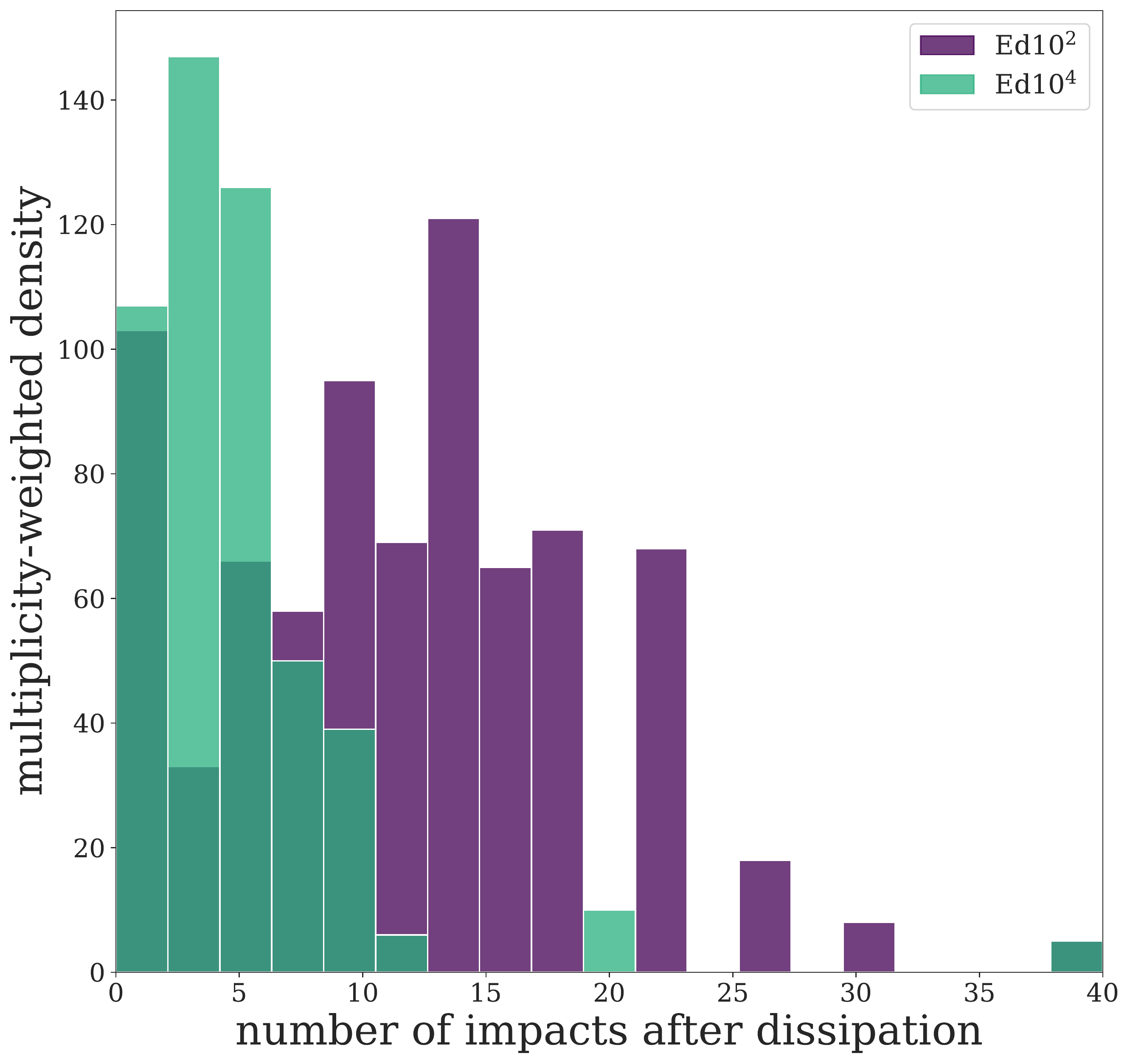}
\centering
\caption{The number of giant impacts during the planet assembly process, per star, after gas dissipation. This distribution has been  multiplicity weighted. \texttt{Ed$10^{2}$} planets are more likely to have suffered an atmosphere removing impact.}
\label{fig:numberofimpacts}
\end{figure}

 In our simulations (intial N=123-220 and particle mass ~0.01$M_{\oplus}$-~1$M_{\oplus}$), the typical impactor radius is high. When we consider all impacts that take place after the gas disk has dissipated, we find that the mass of all impactors in our simulations \textit{exceed} this cutoff mass. Therefore, a simplifed boolean scenario describes our simulations: either an impact \textit{has} occurred and the atmosphere has been boiled off of the core, or an impact \textit{has not} occurred and the primordial atmosphere is retained. We make an additional simplifying assumption: if a giant impact occurs before the gas disk has dissipated, we assume that the planet re-accretes the primordial atmosphere. It is only if the giant impact occurs after gas dissipation (\textless 1 Myr) in the disk that we assume the primordial atmosphere is lost. In Figure \ref{fig:numberofimpacts}, we show a histogram of the number of these post-dissipation impacts that occur in each simulation (that is, the number that occur per disk). 

\subsubsection{Loss Driven by Photoevaporation}
\label{sec:atm_loss_photo}

For photoevaporative atmospheric loss, we turn to the \textsc{atmesc} module of the Virtual Planet Simulator \citep{Barnes20}. \textsc{VPlanet} computes the atmospheric escape of hydrogen-dominated atmospheres in the energy-limited regime while the star is in its pre-main-sequence phase, and transitions to ballistic escape afterward. The XUV luminosity of the star over time is calculated by the \textsc{stellar} module and follows an interpolation of the evolutionary tracks from \citep{Baraffe2015}. Using a Sun analog, we evolved each system forward 5 Gyr with starting envelope mass fractions of 5\%. We consider planets that lose $>$ 90\% of their primordial envelope to be ``stripped," for the sake of simplicity of comparison.
\subsection{Assigning Detection Probability}

Once we have assembled populations of planets, and determined which planets in the simulations retain their primordial atmospheres, we then ``observe" the planetary systems, in order to compare our synthetic planetary sample to \textit{Kepler} observations. We assume a CDPP (Combined Differential Photometric Precision) equal to the median 6-hour rms CDPP for a 12th magnitude star in Kepler Quarter 3 as reported in \citep{Christiansen11}. We consider transiting planets (that is, those with impact parameter $b$ $\le$ 0.9) with a total SNR $>$ 7.1 to be ``detected," where $P$ is the orbital period of the planet, and $\delta$ is the transit depth:

\begin{equation}
    SNR = \sqrt{\frac{\textrm{3 years}}{P}}\frac{\delta}{\textrm{CDPP}.}
\label{eq:cdpp}
\end{equation}

We find that 4.9\% of the full sample of simulated planets are typically detected if giant impacts dominate, and 4.7\% are typically detected if photoevaporation dominates. The slightly lower latter detection rate results from the fact that photoevaporation prevents close-in planets from retaining primordial atmospheres, decreasing their radii and the resulting transit signal. 

\section{Results}
\label{sec:analysis}
With our synthetic sample of observed planets in hand, we now investigate the effect of the dominant atmospheric loss process (whether from photoevaporation or giant impacts) on the properties of the resulting planets. We consider both the effects on the underlying population and the ``observed" population of planets. We find that $\sim 85\%$ of planets experience the same outcome for both mechanisms, whether the atmosphere is lost or retained. However, the retention of the atmospheres of $\sim 15\%$ of planets are contingent on whether giant impacts or photoevaporation dominates. We first focus in Section \ref{sec:atmosphere_occurrence} on this raw likelihood of primordial atmospheric loss for both mechanisms. We then turn to the $\sim 15\%$ of planets with divergent outcomes in \ref{sec:diverge}. The fates of these planets' atmospheres depend on which mechanism is dominant, and therefore these diagnostic planets highlight the conditions under which the assumption of underlying physics matters most. The different resulting radii shift the planetary positions on the resulting period/radius distribution. In Section \ref{sec:intrinstic_radius}, we investigate this radius distribution alone, to characterize the predicted position and relative emptiness of the ``radius gap" \citep{Fulton11}.

\subsection{Raw occurrence of primordial atmospheres}
\label{sec:atmosphere_occurrence}

As we describe in Section \ref{sec:atm_loss_impact}, the masses of all impactors in our simulation exceed the ``cutoff mass" for total atmospheric loss \citep{Biersteker2019}. When we consider a giant impact, therefore, either the primordial atmospheric is left intact, or it is completely lost. We therefore characterize the likelihood of atmospheric loss using a binomial likelihood, where some fraction $f$ of atmospheres are lost to giant impacts. To compare atmosphere loss occurrence due to giant impacts to the more gradual loss driven by photoevaporation, we make a simplifying assumption. If  an amount \textless 10\% of the initial primordial atmosphere remains after 5 Gyr of photoevaporation have elapsed, we consider the atmosphere to be ``lost". This simplification has the advantage of being approximately true and allowing us to again employ a binomial likelihood function for easy comparison to the giant impacts sample. Assuming a uniform prior on the fraction $f$ of planets that lose their primoridal atmospheres, we evaluate the posterior distributions on $f$ for planets resulting from the two sets of initial conditions, whether dynamically cool (\texttt{Ed$10^{2}$}) or dynamically hot (\texttt{Ed$10^{4}$} planets). We find that dynamically cool and dynamically hotter planetary systems retain their atmospheres at similar rates, per Table \ref{tbl:underlying_rate}. However, the loss process itself does affect the atmospheric retention rate: the rate of atmospheric retention is $f=0.52^{+0.02}_{-0.02}$ for giant impacts and $f=0.35^{+0.02}_{-0.02}$ for photoevaporation. 

\begin{table}[]
\centering
\caption{Posterior probability of a planet having an atmosphere for underlying sample}
\label{tab:table1}
\begin{tabular}{lll}
\multicolumn{1}{l|}{}                                     & \multicolumn{1}{l|}{$f_P$}                       & $f_{GI}$                      \\ \hline
\multicolumn{1}{l|}{Ed$10^{2}$} & \multicolumn{1}{l|}{$0.71^{+0.02}_{-0.02}$} & $0.65^{+0.02}_{-0.02}$ \\ \hline
\multicolumn{1}{l|}{Ed$10^{4}$} & \multicolumn{1}{l|}{$0.84^{+0.02}_{-0.02}$} & $0.61^{+0.02}_{-0.02}$  \\ \hline
                                                          &                                                  &                            
\end{tabular}
\label{tbl:underlying_rate}
\end{table}

\subsection{Cases with Divergent Atmospheric Outcomes}
\label{sec:diverge}

There exist a subset of planets whose atmospheric retention depends upon whether photoevaporation or giant impacts are the dominant loss mechanism. While temperature and surface gravity are critical for whether photoevaporative atmosphere loss occurs, loss due to giant impacts is necessarily stochastic in nature. There are therefore a sample of planets that retain atmospheres under one set of assumptions, and lose them under the alternative set of assumptions. With this subset of planets, we can identify the most sensitive parts of parameter space to assumptions about atmospheric loss. 

\begin{figure}[t]
\includegraphics[width=16cm]{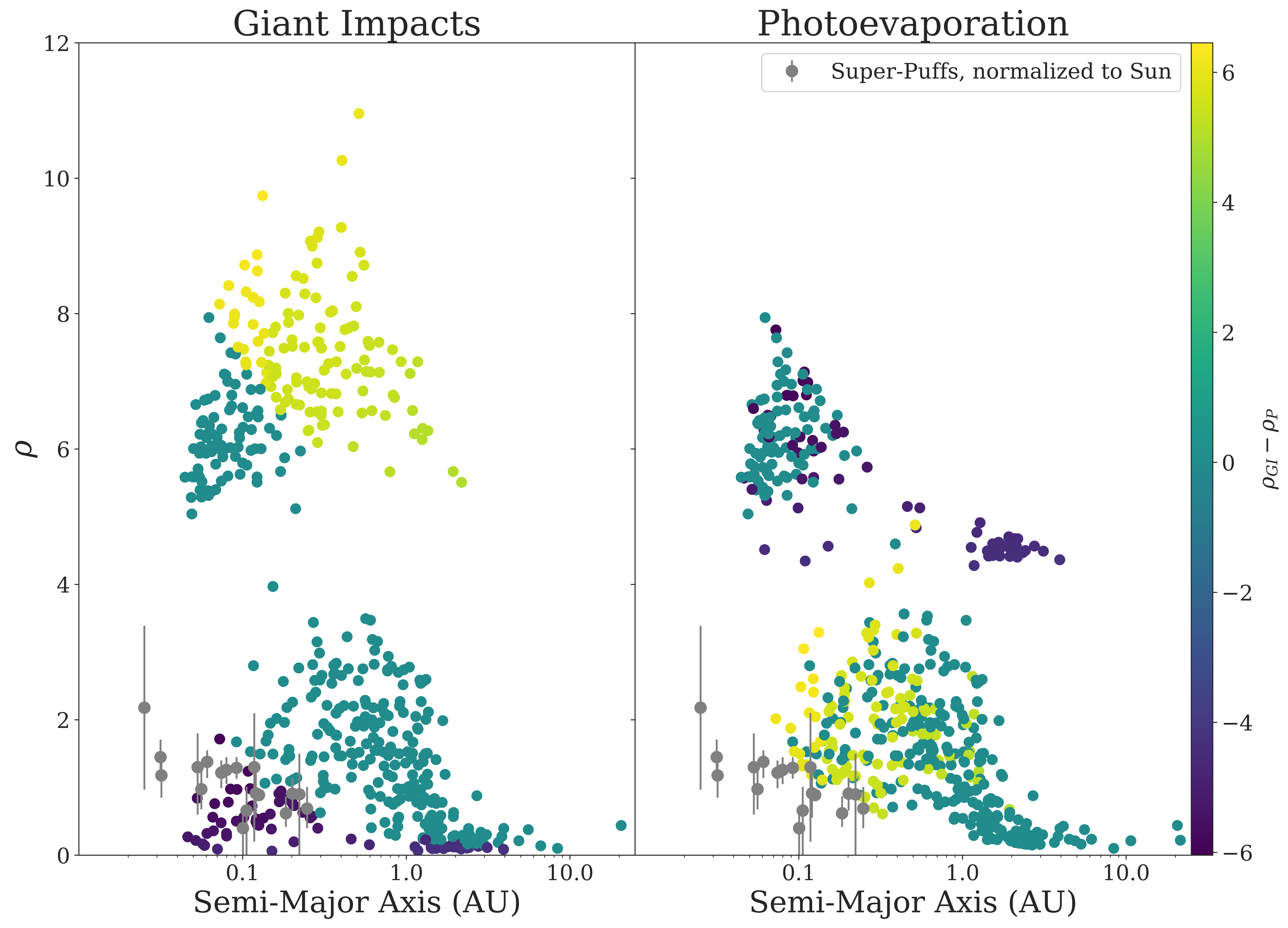}
\centering
\caption{Differing density outcomes for the giant-impact model (at left) versus the photoevaporative model (at right), as a function of semimajor axis. Color-coding corresponds to $\Delta \rho$ between $\rho_{\textrm{GI}}$ and $\rho_{\textrm{Photo}}$. Teal symbols correspond to planets with identical outcomes, or $\Delta \rho\sim$ 0. Purple points correspond to planets with low densities when giant impacts dominate, and high densities when photoevaporation dominates. Planets colored yellow have high densities when giant impacts dominate and low densities when photoevaporation dominates. In grey are known hot exoplanets with low densities, with their semimajor axis position normalized to solar insolation. While they are easily explained if giant impacts dominate (at left), they ought not to exist if only photoevaporative loss always occurs.
}
\label{fig:difference}
\end{figure}

In Figure \ref{fig:difference}, we show how these populations of planets with divergent outcomes shift in density-period space. Considering first giant impacts, we find that populations of low-density planets are present at all periods. These include hot, low-mass planets with primordial atmospheres, which become high-density cores under photoevaporation.  cool, High-density remnant cores are also present at all periods in the sample planets sculpted by giant impacts. In contrast, photoevaporated high-density cores reside mostly at close-in orbits. Overall, we identify three populations that are most sensitive to assumptions about atmosphere loss. These three populations lie in distinct temperature/mass regimes, which we designate Populations A, B, \& C for ease. Population A consists of highly-irradiated ($T_{p}\sim\ 1000 K$) low-mass planets, likelier to lose atmospheres under photoevaporation and keep them under giant impacts. The second group, Population B, is higher mass planets with intermediate irradiation ($T_{p}\sim\ 500 K$), which are conversely likelier to keep their atmospheres under photoevaporation but lose them under giant impacts. Finally, there is a third sample of cold, very low-mass planets ($T_{p}<300 K$), designated Population C. With surface gravity weak enough for photoevaporation to strip their atmospheres, they sometimes retain these atmospheres under giant impacts, if they escape an impactor. 

The presence of this first set of highly-irradiated but low-density planets (Population A here, sometimes denoted ``super-puffs" per \citealt{Lee2016}), particularly challenge the photoevaporative model. This is especially true in the case of extremely low-density exoplanets in multi-planet systems where the XUV history of the host star is well-constrained (Kepler-107, TOI-178 \citep{Bonomo2019,Leleu2021}). In Figure \ref{fig:difference}, these correspond to the purple symbols with orbits clustering near 0.1 AU.  We include in Figure \ref{fig:difference} in grey the densities and periods of a subset of observed ``super-puffs" from the literature (with insolation normalized to solar for inclusion on a common axis). We find that the presence of these planets is completely consistent with a scenario in which giant impacts dominate atmospheric loss, at left; considering only photoevaporative loss, these planets should not exist. Less detectable from transit surveys are the diagnostic Populations B and C of higher-mass stripped cores at temperatures between 1000 and 300 K (present for giant impacts), and cold, small cores (present for photoevaporation), respectively. 

\begin{figure}
\centering

\includegraphics[width=11.5cm]{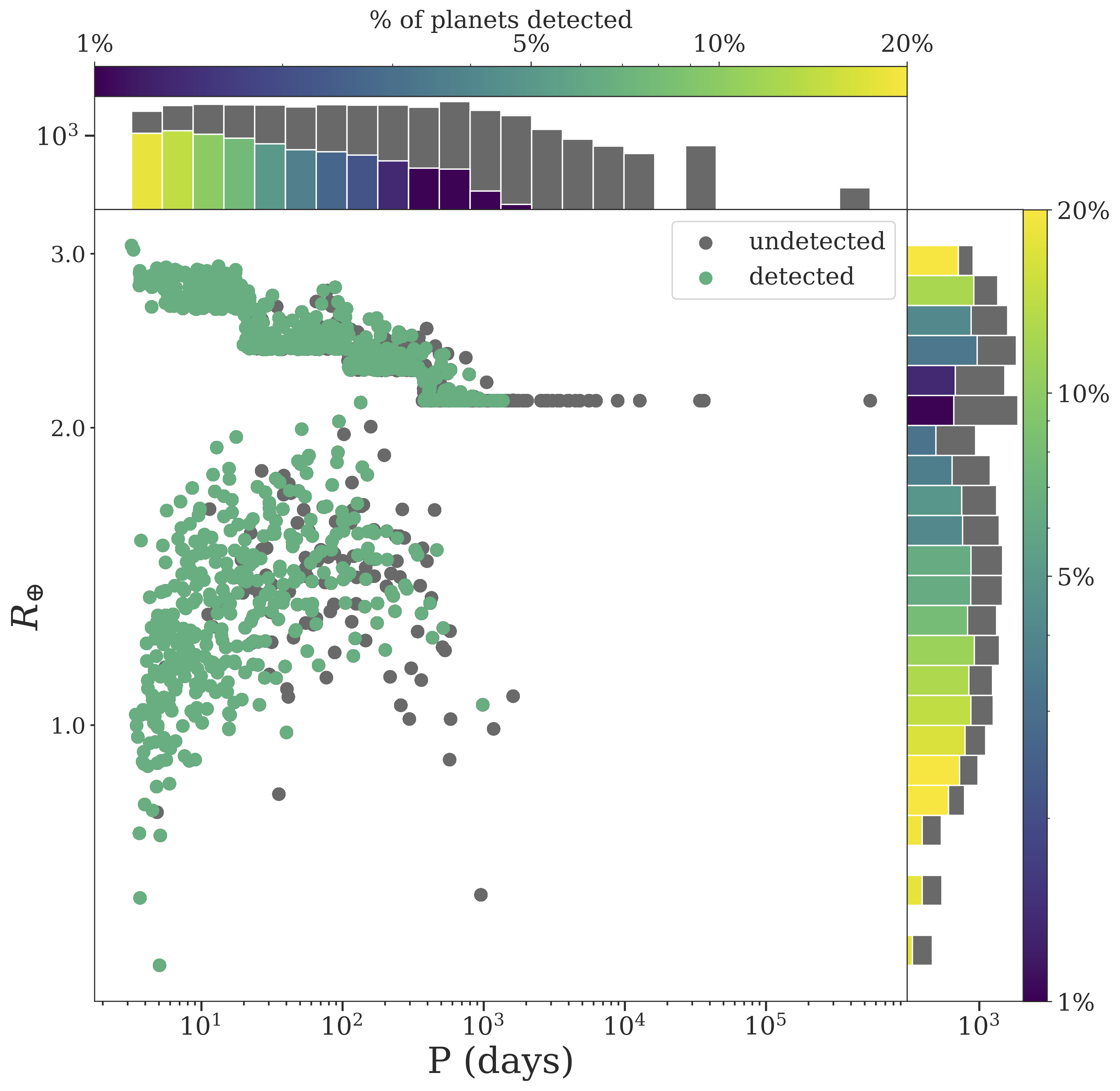}
\label{fullP}

\includegraphics[width=11.5cm]{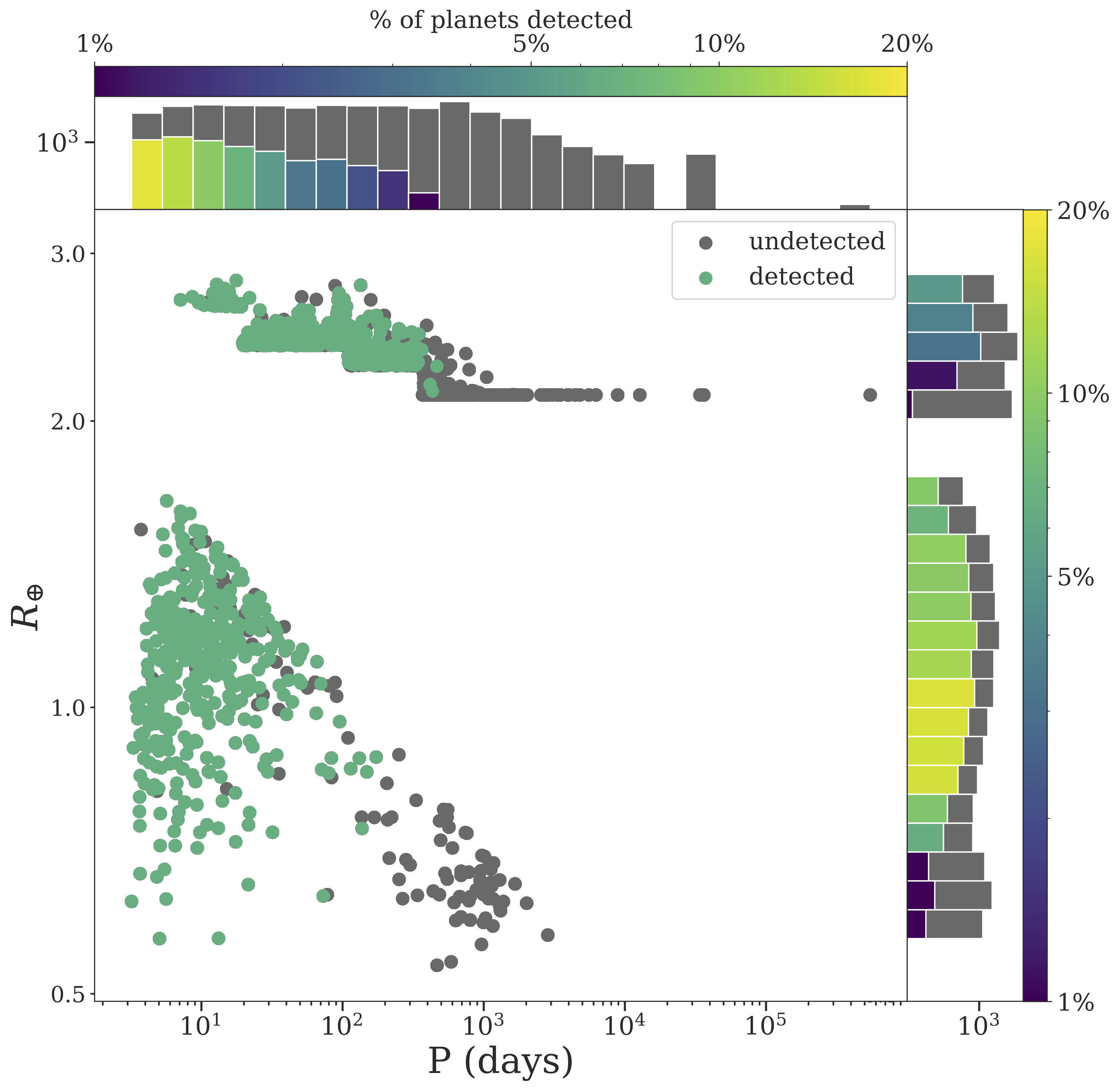}
\label{fig:P_stats}
\vspace{0mm}
\caption{Period-radius distributions for full simulated planet population (grey) and detected sample (green). The two panels show the population sculpted by giant impact atmosphere erosion (left) and photoevaporative erosion (right). The top marginal histogram shows the period distribution of the full sample of planets (grey) and the detected sample (shaded by fraction of planets detected).}
\label{fig:pressure}
\end{figure}

In Figure \ref{fig:pressure}, we show the resulting period-radius distributions for both the underlying and subsequently ``detected" populations of planets. At left, we assume giant impacts are dominant in atmospheric loss at left, and at right, that photoevaporation is dominant. We can now assess the usefulness of the three diagnostic populations most sensitive to loss mechanism, based upon their likelihood of detection.

\begin{figure}
\centering
\includegraphics[width=16cm]{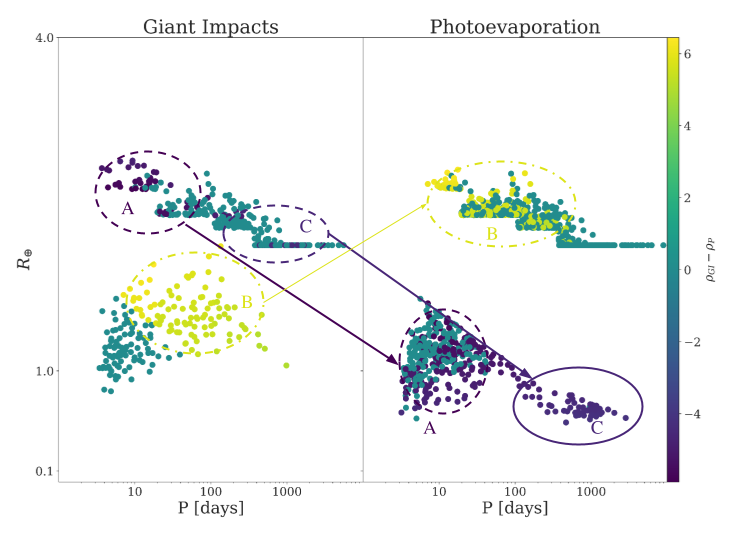}
\label{fig:ABC}
\caption{Period-radius distributions for full simulated planet population, color-coded by the difference in bulk density between giant impacts and photoevaporation.}
\end{figure}

Any difference between these two radius-period diagrams will result from the same set of planets with divergent outcomes, as shown in Figure \ref{fig:difference}. The same three clouds of points in this parameter space, that are apparent in density/semimajor axis space, are most sensitive to atmospheric loss mechanism. We employ this same color scheme (depicting the difference in the bulk density of the planet) in Figure \ref{fig:ABC}, circling each nexus of points and showing their changed location in radius-period space. The hot, low-mass planets of Population A (``super puffs"), that only exist if giant impacts dominate (purple points with $P<50$ days in Figure \ref{fig:ABC}), are apparent in the upper left corner of the giant impacts period-radius diagram. Comparing to the detection rates in that part of parameter space in the top panels of Figure \ref{fig:pressure}, we see that they are nearly always detected with \textit{and} without atmospheres, making them the most useful diagnostic population. Population B are points that are yellow in Figure \ref{fig:difference}; they have retained their atmospheres if we only consider photoevaporation, but lose them if giant impacts are most important. These are typically higher-mass planets with temperatures near 500 K. These  planets, when placed on the giant impacts period-radius diagram shown in the left panel of Figure \ref{fig:pressure}, comprise the stripped cores with mean size 1.5 $R_{\oplus}$ and periods \textless 10 days. They are drawn approximately uniformly in log period from among the upper cloud of sub-Neptunes (reflecting the stochastic nature of giant impacts). As stripped cores if giant impacts dominate, they are detected with approximately 50\% probability. Finally, there exist the Population C planets: tiny and cool stripped cores if photoevaporation dominates (with mean $R_{p}$ of 0.5 $R_{\oplus}$ and period of 1000 days, in the right panel of Figure \ref{fig:pressure}). Under giant impacts, these small planets mostly retain their atmospheres, but have low enough surface gravity that photoevaporation is sufficient to remove them. They make up the bulk of stripped planets with periods longer than 10 days in the photoevaporation-dominant scenario, and are very diagnostic in the sense that they should \textit{not} exist if only giant impacts are considered. However, they are rarely detected when stripped (indicated by their grey color in Figure \ref{fig:pressure}), and are therefore least useful as an observational diagnostic tool. 
\subsection{Radius distribution}
\label{sec:intrinstic_radius}

To understand the effect of loss mechanism on the observed radius distribution, we first examine the intrinsic underlying distribution, before turning to the ``observed" distribution. We find key differences in the radius distribution of planets dependent upon whether giant impacts or photoevaporation is dominant, due almost entirely to the fate of planets designated as Population B in Figure \ref{fig:ABC}. 

\begin{figure}
\centering
\includegraphics[width=16cm]{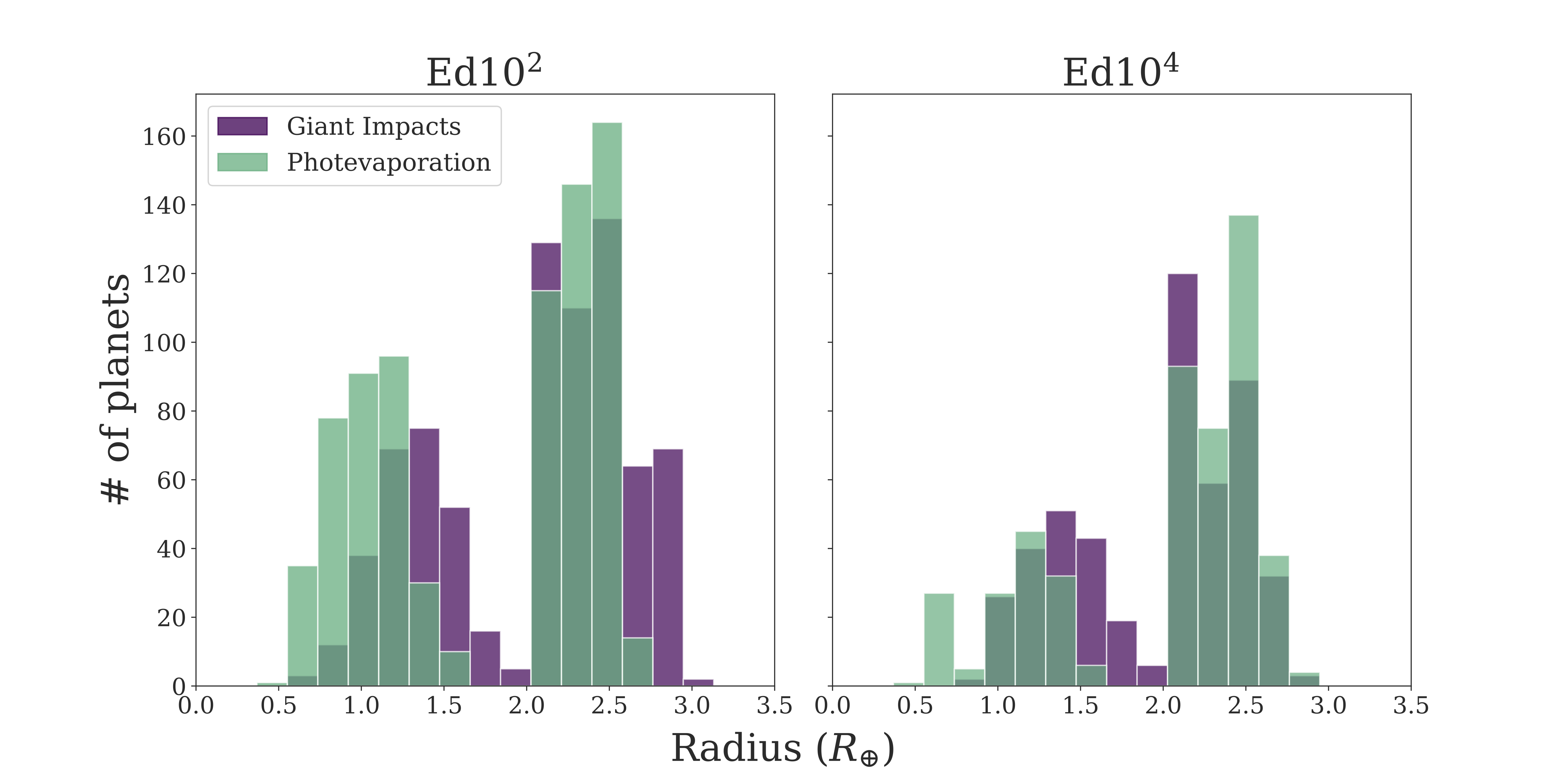}
\caption{The intrinsic radii distributions of \texttt{Ed$10^{2}$} (associated with dynamically cooler outcomes) and \texttt{Ed$10^{4}$} (associated with dynamically hotter outcomes) simulations.}
\label{fig:lesser}
\end{figure}

We first examine the effect of loss mechanism, between the samples of cooler and warmer dynamical temperature. Because these two ensembles, \texttt{Ed$10^{2}$} and \texttt{Ed$10^{4}$}, produce different mass and Hill spacing distributions, it stands to reason that they may be differently impacted by atmospheric loss mechanism. In Figure \ref{fig:lesser}, we show the histograms of resulting planet radii for both modes of planet formation. The mode that produces the dynamically cooler systems of more coplanar planets ((\texttt{Ed$10^{2}$}) is shown at left, and the mode that produces dynamically hotter systems with less planets (\texttt{Ed$10^{4}$}, at right). 

  \texttt{Ed$10^{2}$} generates a larger number cores of smaller average mass. Small cores that retain atmospheres tend to cluster near $2 M_{\Earth}$. \texttt{Ed$10^{4}$} generates fewer, but larger cores. The presence of larger cores in the latter sample appears in the relative shift to higher radii in both modes of the radius distribution. The effect of higher mass planets in \texttt{Ed$10^{4}$} is also to compress the atmospheres of planets with large radii. The low mass ($< .5 M_{\Earth}$) planets of \texttt{Ed$10^{2}$} generate the ``super-puff" \citep{Lee2016} planets. With high insolation and low mass, they retain their extended atmospheres only if we do not consider photoevaporation.

\begin{figure}[h!][width=16cm]
\centering
\includegraphics[width=8in]{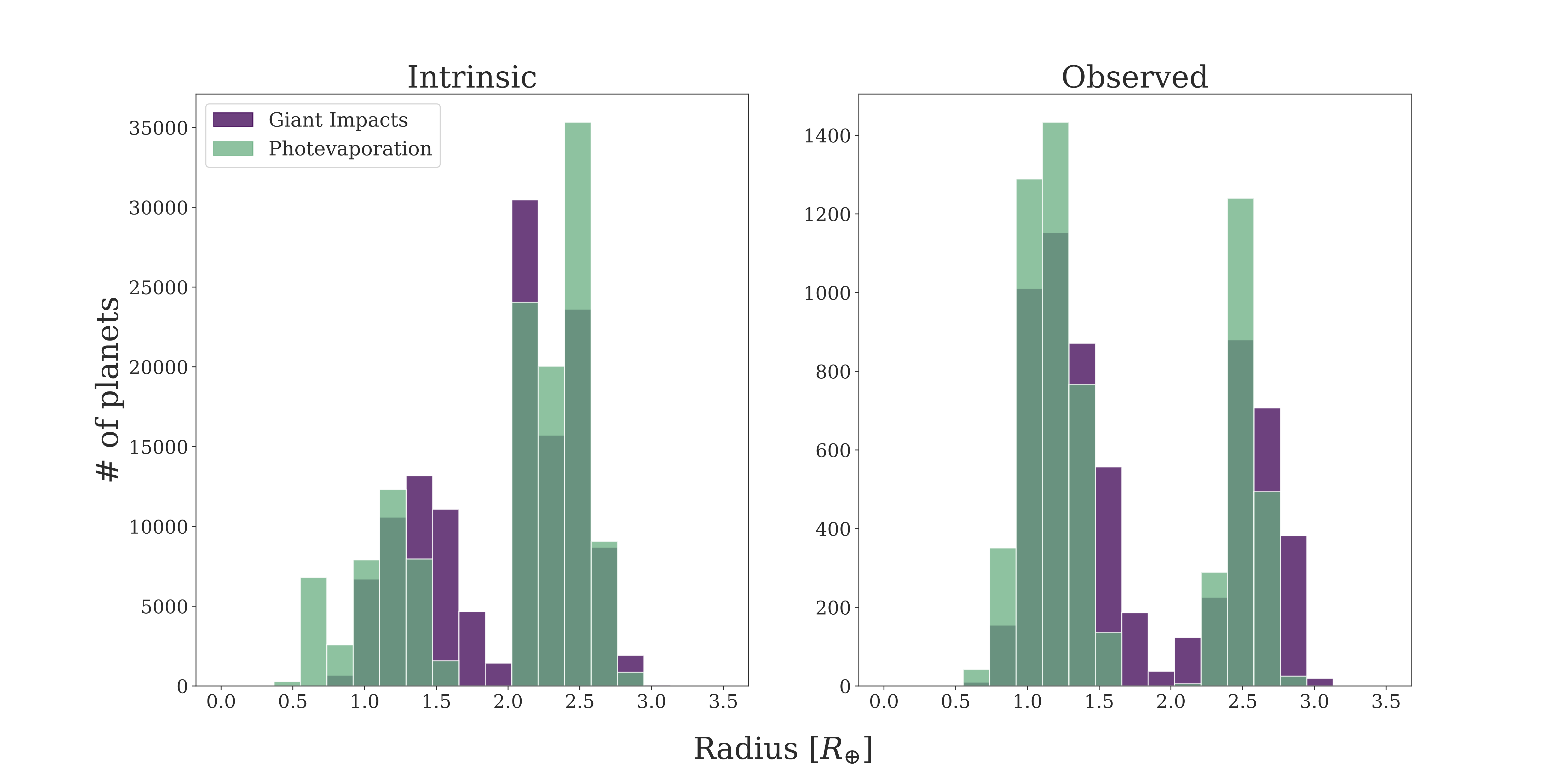}
\caption{The intrinsic versus ``observed" radius distributions of exoplanets, when employing a mixture model of \texttt{Ed$10^{2}$}  and \texttt{Ed$10^{4}$} simulations to mimic Kepler planet occurrence per Section \ref{sec:assembly}}
\label{fig:intrinsic}
\end{figure}

We turn in Figure \ref{fig:intrinsic} to the resulting \textit{observed} distributions of planetary radii. These samples are now drawn from the mixture models described in Section \ref{sec:assembly}, with 7\% from the {Ed$10^{2}$} simulations and 93\% from the {Ed$10^{4}$} simulations, and we have retained only the ``detected" planets. Since the transit detection method selects strongly for short periods, the observed planet populations are mostly drawn from this close-in population. In bother cases, this translates to a higher \textit{observed} fraction of stripped cores, compared to the intrinsic underlying occurrence (that is, the observed relative heights of the two radius peaks is reversed from their true relative heights). In both cases, this is due to the higher incidence of mini-Neptunes (with retained H/He atmospheres) at longer orbital periods. 

\section{Discussion: Observational consequences}
We identify four potential observational diagnostics, useful for distinguishing between a large population of exoplanets sculpted primarily by photoevaporation and a population sculpted primarily by giant impacts. These are (1) the morphology of the radius gap (Section \ref{sec:gap}), (2) the occurrence of H/He atmospheres as a function of transit multiplicity (Section \ref{sec:multiplicity}), (3) the size-ordering (or lack thereof) in multi-planet systems (Section \ref{sec:monotonicity}), and (4) the ease of distinction between the densities of stripped planets and planets with primordial atmospheres (Section \ref{sec:clustering}).

\subsection{Radius gap}
\label{sec:gap}
 The morphology of the exoplanet radius distribution  is the subject of active investigation \citep{Fulton18,Berger20,VanEylen18}.
 The location of the ``gap" separating smaller super-Earths from larger sub-Neptunes changes with spectral type, and even within the same spectral type, stars vary in XUV flux history. This complicates predictions for the gap morphology from photoevaporation. We find that the two mechanisms we consider in this manuscript, photoevaporation and giant impacts, are distinguishable by the relative emptiness of the radius gap. This is due to the divergent fates of the planets we have called ``Population B", which retain their atmospheres under photoevaporation and lose them under giant impacts. In Figure \ref{fig:P_stats} depicting the radius/period distribution of planets, the radius distribution is projected against the right-hand side of both panels (for both the full and observed samples). Population B accounts entirely for the change in the relative emptiness of the gap. These planets are coded as in yellow in the density difference plot of Figure \ref{fig:difference} and \ref{fig:ABC}, and comprise the largest and highest-density stripped cores. As a group, these planets possess orbital periods $>$10 days and surface gravities large enough to retain atmospheres under photoevaporation. Their atmospheres are lost only when we consider giant impacts. We examine the intrinsic radius distributions (before observation) more closely in \ref{fig:intrinsic}. While the position of the gap at $1.8-1.9 R_{\oplus}$ is unchanged, it is the \textit{relative emptiness} of the gap that depends strongly upon which loss mechanism is dominant. For photoevaporation, the gap is entirely empty, and there exists a clear separation between planets with and without atmospheres. This is because we predict few, if any, high-mass stripped cores under photoevaporation. Under the set of assumptions we have used here, only a giant impact can strip the atmosphere of the larger cores ($\ge$1.7 $R_{\oplus}$ at $P=10$ days, or $\ge$1.2 $R_{\oplus}$ at 100 days).
 
\begin{figure}
\includegraphics[width=16cm]{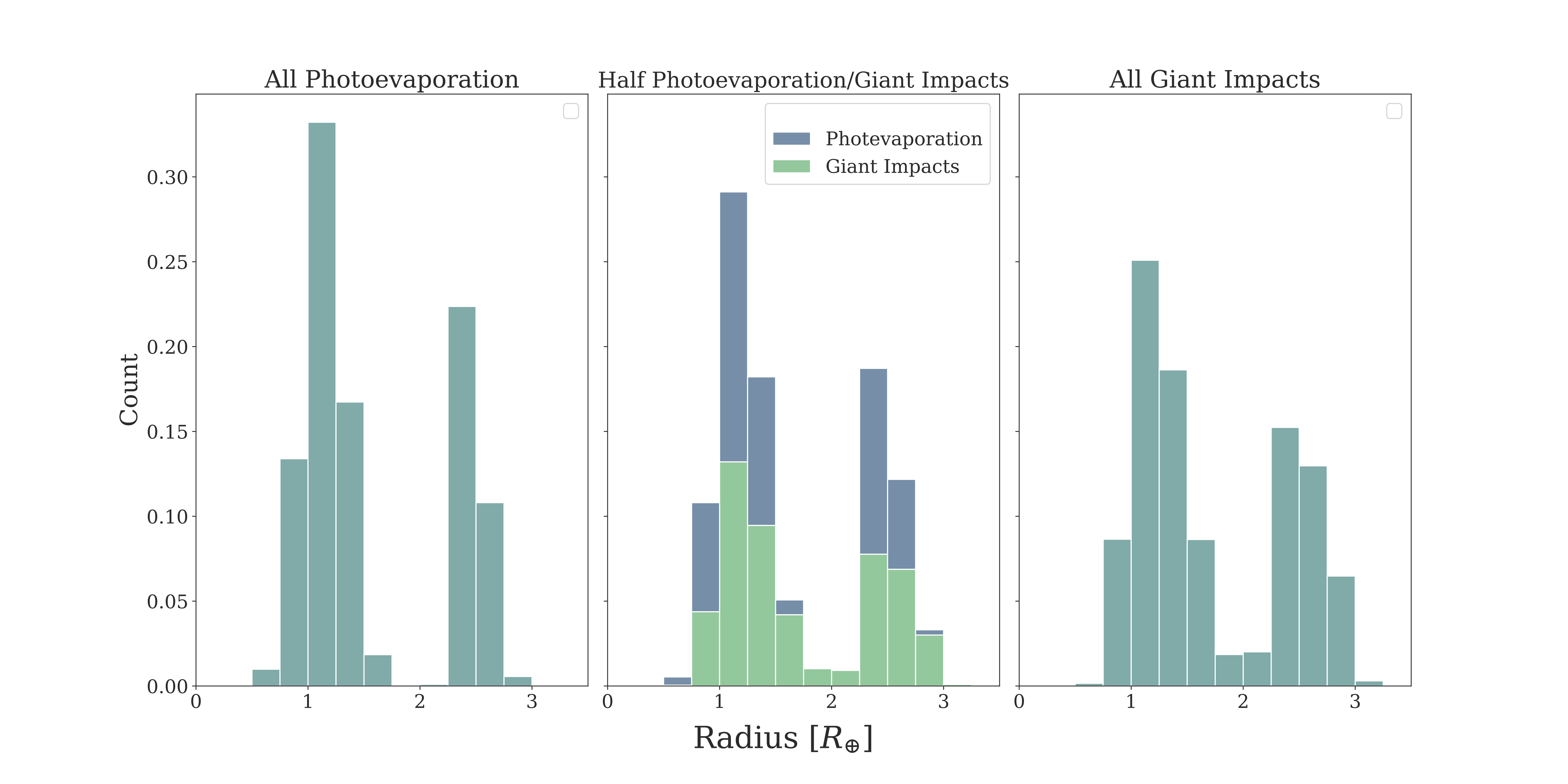}
\centering
\label{fig:comparison}
\caption{A comparison between planet radii distributions in scenarios where photoevaporation is dominant (far left), where giant impacts are dominant (far right), or and where both kinds of atmosphere loss occur in equal proportion (center).
}

\end{figure}

In Figure \ref{fig:comparison}, we show the predicted gap morphology in three scenarios: only photoevaporation, only giant impacts, and a scenario with 50\% likelihood of either outcome. The latter two are associated with the presence of planets within the radius gap, consistent with the \cite{Fulton18} findings that the gap is not empty (that is, that the presence of planets with radii in the gap cannot be explained by radius uncertainty alone). 


\subsection{Occurrence of atmospheres versus planetary multiplicity}
\label{sec:multiplicity}
The ratio of \texttt{Ed$10^{4}$} planets to \texttt{Ed$10^{2}$} planets drops precipitously with increasing transit multiplicity, with we might expect to see different atmospheric outcomes as a function of transit multiplicity. In particular, while singly- and doubly-transiting systems are drawn from a mixture of populations, systems with $\ge3$ transits are drawn exclusively from \texttt{Ed$10^{2}$}. 

Single transit systems are mostly drawn from \texttt{Ed$10^{4}$}, given the larger orbital spacing between planets resulting from these simulations. On average, these planets are more massive than planets in \texttt{Ed$10^{2}$}. However, this depends on orbital period in a similar manner to \cite{Moriarty2016} (see Figure 1 of that work): while \texttt{Ed$10^{4}$} cores grow more massive with increasing orbital period, \texttt{Ed$10^{2}$} cores tend to stay smaller and more uniform.  The increased mass and corresponding increase in surface gravity make \texttt{Ed$10^{4}$} cores likelier to retain atmospheres under photoevaporation, when compared to a \texttt{Ed$10^{2}$} planet receiving the same incident flux. 

\begin{figure}[h!]
\centering
\includegraphics[width=12cm]{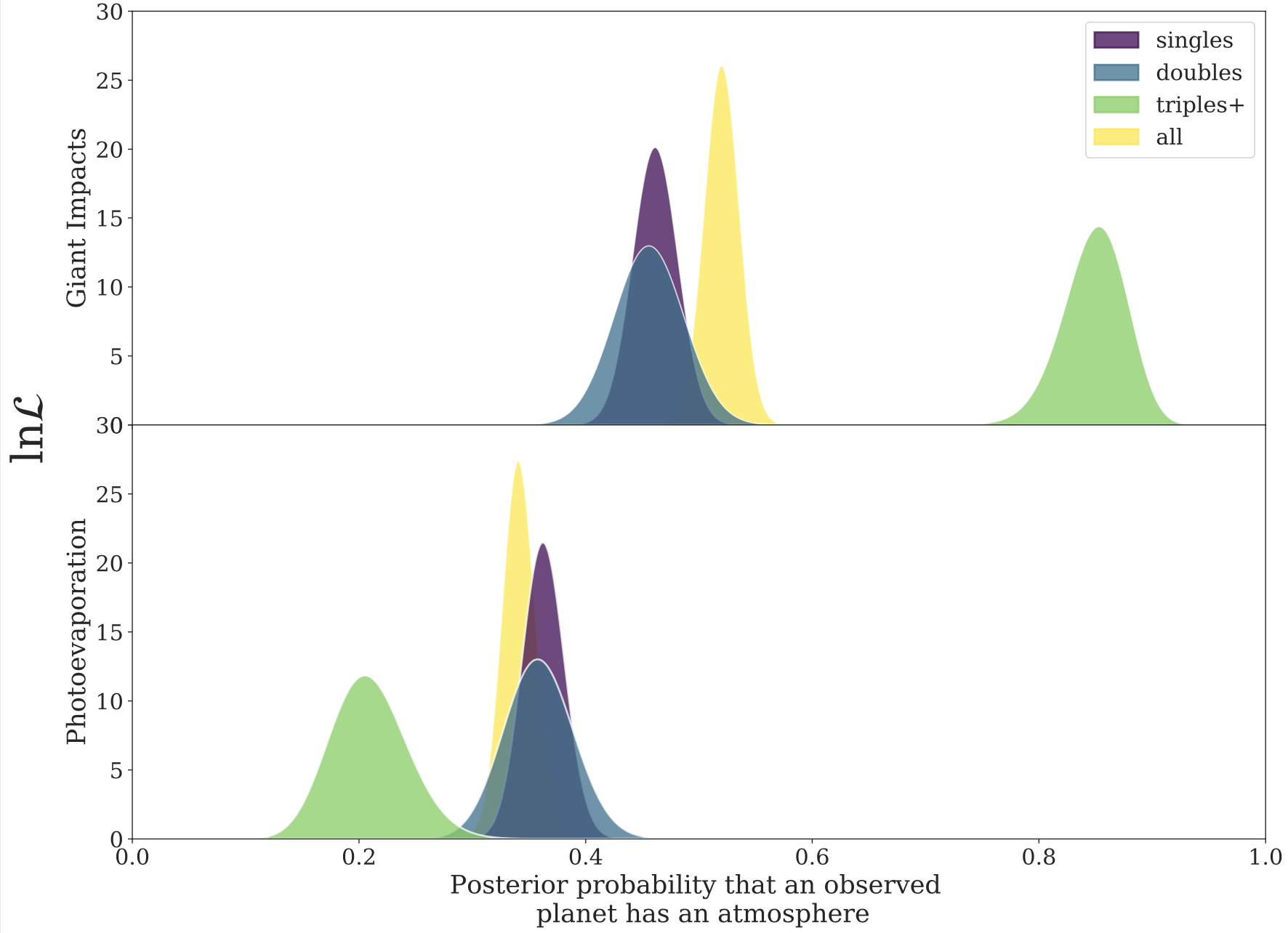}
\caption{The probability the an observed planet retains an atmosphere under both models of atmosphere loss. Planets in STIPS are highly likely to retain their atmospheres when sculpted by giant impacts.}
\label{fig:atmo}
\end{figure}

Figure \ref{fig:atmo} shows the resulting likelihood of atmospheric retention among \textit{transiting} planets as a function of their transit multiplicity, whether giant impacts (upper panel) or photoevaporation (lower panel) is the dominant atmospheric loss mechanism. The yellow posterior distributions corresponds to the binomial likelihood $f$ of atmospheric retention of all planets, transiting and non-transiting (described in Section \ref{sec:atmosphere_occurrence}. Singly- and doubling-transiting systems are drawn from a similar mixture of \texttt{Ed$10^{2}$} and \texttt{Ed$10^{4}$} disks, which is reflected in their identical atmospheric retention rates. The retention rates for single-transit and double-transit systems are also consistent with the overall retention rate.  

It is in the atmospheric retention rate in systems with $\ge3$ transiting planets that we see a marked effect, in two ways. First, atmospheric retention rates among systems with 3 or more transiting systems are not reflective of the total retention rate. This is true for both the photoevaporation-dominant and giant-impacts-dominant scenario. In the case of giant impacts, the likelihood of atmospheric retention is $0.80^{+0.04}_{-0.03}$ among systems with $\ge3$ transits, \textit{higher} than the $0.52\pm0.02$ rate among all systems. In the case of photoevaporation, the likelihood of atmospheric retention is $0.2\pm0.03$ among systems with $\ge3$ transiting planets, \textit{lower} than the underlying 0.35$\pm$0.03 rate among all systems. Secondly, while the atmospheric retention rates among singly- and doubly-transiting systems are similar between giant impacts and photoevaporation (0.46 versus 0.38 for single transit systems, for example), they are very different for systems with $\ge3$ transits: in multi-transit systems under a photoevaporation-dominant scenario, only 1 in 5 detected planets has retained their primordial atmosphere, but 4 in 5 detected planets have retained it if the giant impacts mechanism is dominant. This is directly reflective of the outcome of \texttt{Ed$10^{4}$} simulated disks. Since triply-transiting systems are all drawn from this population (with more total planets in more closely-spaced and more coplanar orbits), they are a direct reflection of the strongly divergent predicted outcomes for STIPs under photoevaporation and giant impacts.   

\begin{table}[]
\centering
\caption{Posterior probability of a planet having an atmosphere for observed sample}
\label{tab:table2}
\begin{tabular}{l|l|l}
        & $f_P$                       & $f_{GI}$                      \\ \hline
singles & $0.38^{+0.03}_{-0.03}$ & $0.46^{+0.03}_{-0.03}$ \\ \hline
doubles & $0.36^{+0.03}_{-0.03}$ & $0.49^{+0.02}_{-0.02}$ \\ \hline
multis  & $0.22^{+0.03}_{-0.03}$ & $0.80^{+0.04}_{-0.03}$

\end{tabular}
\label{tbl:retain_multi}
\end{table}

\subsection{Monotonicity}
\label{sec:monotonicity}
 STIPS exhibit similarity between neighboring planets, with evidence for monotonically increasing planet size with period (the ``peas-in-a-pod" phenomenon, explored in detail in \citealt{Millholland17,Weiss_2018,Weiss_2020,Murchikova_2020,Millholland21} and others). The typical ratio in size of super-Earths to neighboring sub-Neptunes ($R_{\textrm{SN}} \approx 1.7 R_{\textrm{SE}}$, per \citealt{Millholland21}) is consistent with multiple models for the loss of primordial atmospheres, including core-powered mass loss (e.g. \citealt{Gupta2019}) and photoevaporation. The relationship between planetary size and orbital period within the same system presents another diagnostic \citep{Ciardi2013,Kipping18,Millholland17,Weiss_2018}. An outstanding observational challenge to the photevaporation hypothesis is the existence of neighboring planets with divergent densities. Such systems include TOI-178 \textit{d} and \textit{e} \citep{Leleu2021} and Kepler-107 \textit{b} and \textit{c}. We explore the intra-system patterns associated with giant impacts and photoevaporation here. 
 
 Giant impacts, given the stochastic nature of impactors, provides a natural possible explanation for the existence of dissimilar neighbors. Monotonic increase in planet size is consistent with photevaporative loss of primordial planet atmospheres. Closer planets are more irradiated and therefore likelier to be evaporated cores. A hypothesized giant-impacts-domninant scenario, in contrast, has competing effects in favor and against a monotonic trend. In one sense, we would expect a giant-impacts-dominant scenario to result in the same observed smaller-to-larger size ordering as photoevaporation, since the shorter dynamical timescale of the inner systems allow for more orbit crossing events. This would results in a modest increase in probability of an impact for close-in planets. On the other hand, the stochastic nature of giant impacts could allow for a less ordered scenario. Populations A and B in the giant impacts scenario described in Section \ref{sec:diverge} represent close-in sub-Neptunes (with retained atmospheres) and modestly irradiated cores, neither of which exist in the photoevaporation dominant scenario. 
 
 To assess the predicted effect of giant-impact-dominated atmospheric loss on size ordering, we employ the monotonicity statistic as defined in \cite{Gilbert2020}: 
 
 \begin{equation}
 \mathcal{M}=\rho_{S}\mathcal{Q}^{1/N}.
 \end{equation}
 
 where $\rho_{S}$ is the Spearman rank-order coefficient, calculated from the planet masses. This coefficient can be 1 (for perfectly positive monotonic systems), −1 for perfectly negative monotonic systems, 0 for systems with no evidence of monotonic behavior. We calculate $\mathcal{M}$ for each of our systems with observed and detected planets. For additional description of $\mathcal{M}$, including the power dependence of the mass partition function $\mathcal{Q}$, we refer the reader to \cite{Gilbert2020}. 
 
\begin{figure}[h!]
\centering
\includegraphics[width=12cm]{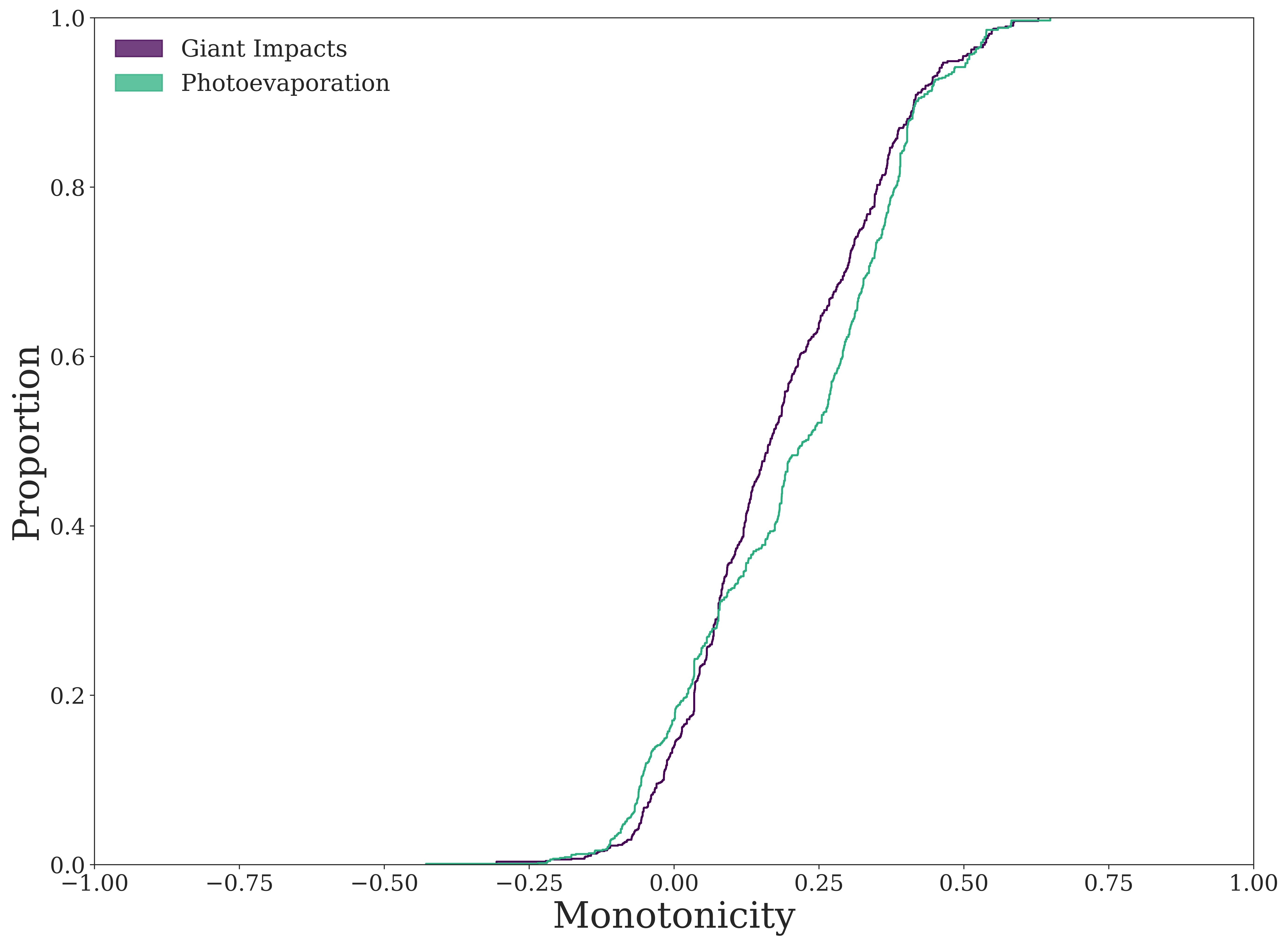}
\caption{The distribution of observed simulated system monotonicity as defined in \cite{Gilbert2020}. Systems close to 1 are more ordered in radius from smallest to largest; systems close to -1 are ordered from largest to smallest. Giant impacts and photoevaporation largely follow the same trend of increasing size ordering, but systems sculpted by giant impacts are ordered to a lesser degree. 
}
\label{fig:monotone}
\end{figure}

We show the resulting cumulative distributions for $\mathcal{M}$ in Figure \ref{fig:monotone}, where systems closer to 1 exhibit more monotonicity from smallest to largest. We find that the giant-impacts-dominant scenario is associated with modestly lower monotonicity coefficients: taking $\mathcal{M}$=0.25 as a sample value, we find that 50\% of photoevaporated systems have $\mathcal{M}\ge$0.25 (that is, more highly ordered), whereas this fraction is 35\% for giant-impacts dominated scenarios. \cite{Gilbert2020} calculated the monotonicity coefficient for the observed sample of \textit{Kepler} planets as well. With 25\% of systems exhibiting $\mathcal{M}\ge$0.25 in that sample, it more closely resembles our giant-impacts distribution distribution, but a detailed comparison of monotonicity distributions to the observed data requires additional study.
 
\subsection{Density clustering}
\label{sec:clustering}
 Our final diagnostic for differentiating between giant-impacts or photoevaporation-dominant scenarios employs planetary bulk density. The populations of (1) planets with lost atmospheres and (2) planets with retained atmospheres hew to the approximate bulk densities of rock (at $\sim$6 g/cm$^{-3}$) and water (at $\sim$2 g/cm$^{-3}$), respectively, as shown in Figure \ref{fig:difference}. However, the distance between the two modes of these bimodal density distribution depends on whether photoevaporation or giant impacts is dominant. The two modes are further separated from one another when we consider only giant impacts. This is because a giant-impacts-dominant scenario allows for extremely high \textit{and} extremely low density planets to exist. These planets are not predicted to exist when photoevaporation is dominant, by the following reasoning. On the high-density end, giant impacts allow for high-mass, high-density stripped cores (these are the yellow ``Population B" planets in Figures \ref{fig:difference} and \ref{fig:ABC}). Under photoevaporation, the surface gravity of these cores is high enough that their atmospheres are not lost even under heavy insolation. On the low-mass end, giant impacts allow for low-mass planets to retain their primordial atmospheres, thereby allowing for small planets with very extended atmospheres (these are the close-in, purple ``Population A" planets in Figures \ref{fig:difference} and \ref{fig:ABC}). These atmospheres are readily lost under a photoevaporation scenario. The existence of these extremely high and low density planets means that if giant impacts dominate, the resulting density distribution of planets will be more easily resolved into the two distinct modes. 
 
 

\begin{figure}
\centering
\includegraphics[width=16cm]{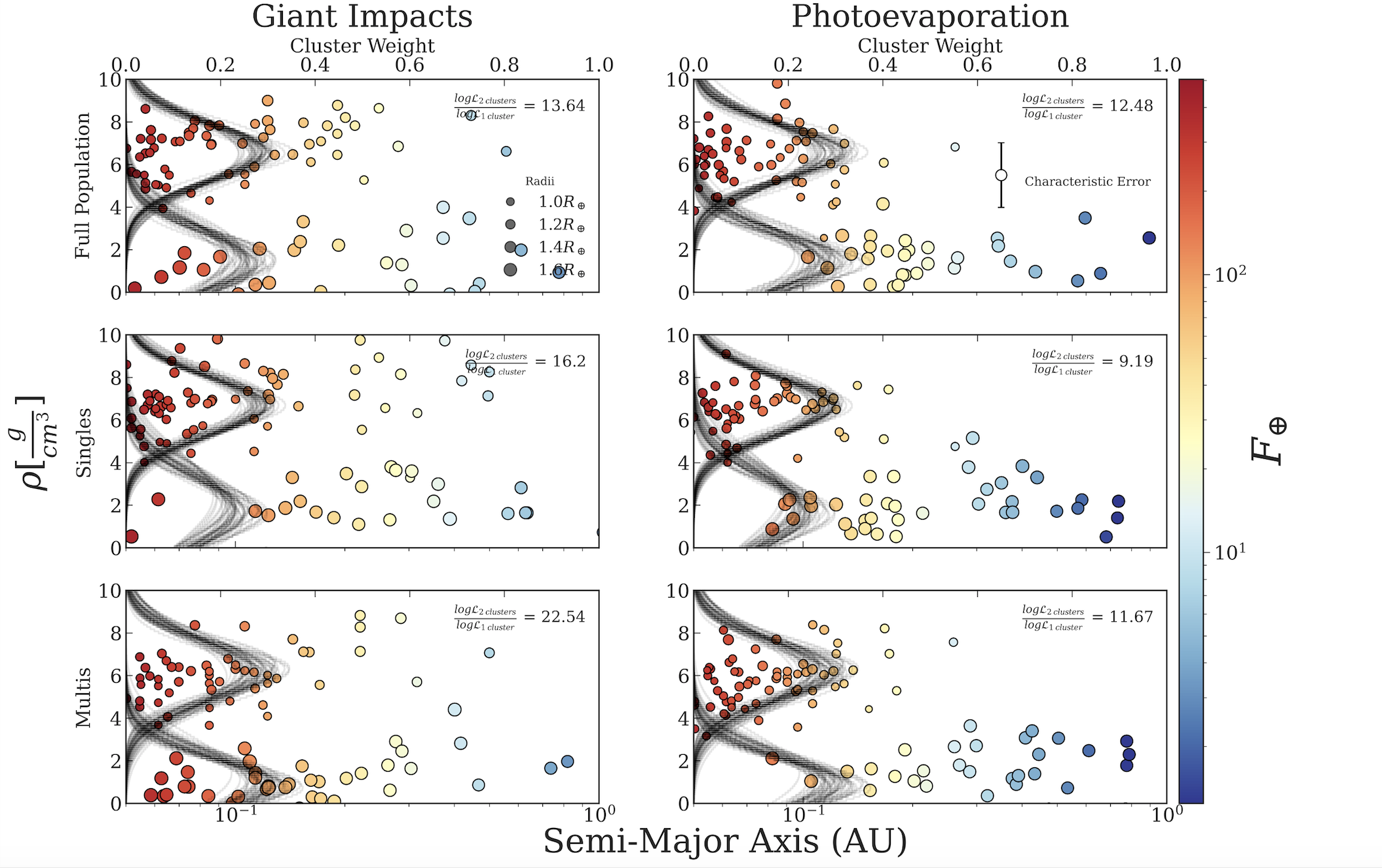}
\caption{The density vs.semi-major-axis distribution of detected planets for both models of atmosphere erosion. The more pronounced clustering of planets sculpted by giant impacts makes it possible to distinguish which model is the dominant one.}
\label{fig:clustering}
\end{figure}

For this reason, the dominant mode of atmosphere loss is encoded in how easily we can distinguish the Earth-density and Neptune-density modes from one another, for a given density uncertainty. The existence of both higher \textit{and} lower density planets under a giant impacts-dominant model makes it easier to resolve one mode from the other. To assess the diagnostic usefulness of this feature, we draw 100 samples of 100 planets each from our observed populations, to which we add a mean density error of 1 g cm$^{-3}$ (characteristic of, for example, a radius uncertainty of 3\% and a mass uncertainty of 8\% for a 5 $M_{\oplus}$ planet with density of 8 g/cm$^{-3}$, per \citealt{Malavolta18}). We then take the ratio of the likelihoods for two hypothetical models: a unimodal density distribution and a bimodal one. These comprise a set of nested models, with the unimodal distribution being merely a special case of the bimodal distribution (with the amplitude of one mode set to zero). For this reason, we can interpret the log of this likelihood ratio as relative preference between the uni- and bi-modal models, given an uninformative prior (c.f. \citealt{Feroz08}).  In \ref{fig:clustering}, we show a representative sample of 100 planets along with 100 representative bimodal model samples. 
 
    We find that, with a typical density uncertainty of 1 g cm$^{-3}$, a giant impacts-dominated sample emerges as bimodal with roughly twice the confidence (that the observed density distribution is described by two Gaussians rather than one) of the same sample of planets subjected to only photoevaporation. This translates to about half as many planets being required to statistically distinguish the two density modes, if giant impacts are the dominant atmospheric loss mechanism. Figure \ref{fig:clustering} highlights this finding, with the ratio of likelihood between the bimodal and unimodal distributions in the upper right of each panel. We repeated the experiment with a typical uncertainty of 2 g cm$^{-3}$, but even with 100 mass measurements, we could not meaningfully distinguish between the two density modes.
    
    We find that this density diagnostic is most effective when considering multi-transiting systems. This is because multi-transiting systems are likelier to contain planets from ``Population A", which that are strongly diagnostic of their atmospheric loss mechanism. These have very low densities under giant impacts and high densities under photoevaporation (see Figure \ref{fig:ABC}). Fortuitously, obtaining density measurements for planets in multi-transiting systems is made easier by the possibility of mass measurement with the transiting timing variation method \citep{Holman05, Agol05}.

 \section{Conclusions}
 
 We conclude that the two atmospheric loss mechanisms we consider here should produce divergent demographic-scale effects, due to the 15\% of planets whose atmospheric retention depends on which loss mechanism is dominant. The predicted differences expected from these two atmospheric loss processes are hypothetically observable, from our analysis of a simulated Kepler-sized sample of exoplanets. We have employed a suite of late-stage planet formation simulations published by \cite{Dawson16} to craft this population; these were already shown in that work to reproduce important \textit{Kepler} observables such as transit multiplicity and period ratio between adjacent planets. By tracking impactors after the nominal gas dissipation phase at 1 Myr, we have modeled the effects of impact-driven atmospheric loss on a large population of planets. By then modeling photoevaporative atmospheric loss in the same set of planets, we can directly compare the fingerprints of these two processes on a common sample. 
Similar to the findings of  \cite{MacDonald2020}, we conclude that these two processes form planets with different radii largely depending on when planets complete their assembly- before or after the gas disk dissipates. Planets the finish forming after the gas disk dissipates run the risk of having their atmosphere stripped by an impact, regardless of separation from the host star. Systems that finish assembly before the gas disk dissipates lose atmosphere to photevaporation in order of distance from the host star, owing to the "peas-in-a-pod" phenomenon in planet masses. 
 The 15\% of planets that experience divergent outcomes, dependent on which loss mechanism is dominant, fall into three main categories. We have designated these as Populations A, B, and C. Populations A and C are predicted to retain H/He atmospheres when giant impacts are dominant, and to lose them if photoevaporation is dominant. Population B, in contrast, is comprised of planets that are predicted to retain H/He atmospheres if photoevaporation is dominant, and to lose them only when giant impacts play an important role. The predicted detection rates of planets in these populations enable us to identify which of these diagnostic planets are most useful from an observational standpoint. At a demographic level, we identify four diagnostic tools for discerning which atmospheric loss mechanism is dominant. These are:
 
 \begin{itemize}
     \item The relative emptiness of the radius gap, which is completely empty when planets are subjected to photoevaporation, but not empty when giant impacts are dominant,
     \item The occurrence of primordial atmospheres as a function of transit multiplicity, where the observed number of transiting planets retention under giant impacts is correlated with the likelihood of atmospheric retention,
     \item The size-ordering of planets as orbital period increases (also denoted ``monotonicity''), where giant impacts produce more sets of neighboring planets with very different densities, and
     \item The existence of very high and low density outliers when giant impacts are dominant. This results in an ease of distinguishing between rocky stripped cores and sub-Neptunes at a population level with greater ease. 
 \end{itemize}
 
  We have not considered the core-powered mass loss scenario here, whereby primordial atmospheres are lost due to the luminosity of the cooling rocky core \citep{Ginzburg2018}. We leave a comparison between this and other hypothetical atmospheric erosion mechanisms for future work. Future observations to refine the resolution of the radius gap and radial velocity campaigns for a few hundred planets will be sufficient to determine which physics is dominant for atmospheric loss in the latest stages of planet formation.

\bibliography{qac_refs.bib}

\end{document}